\documentclass[12pt]{article}
\usepackage{amsmath}
\usepackage{graphicx}
\usepackage{url} 
\usepackage{amssymb}
\usepackage{algorithm}
\usepackage[noend]{algpseudocode}
\usepackage{float}
\usepackage{adjustbox}
\makeatletter
\def\algbackskip{\hskip-\ALG@thistlm}
\makeatother
\usepackage{natbib}
\usepackage{multirow}
\newcommand{\blind}{0}

\begin{document}

\def\spacingset#1{\renewcommand{\baselinestretch}%
{#1}\small\normalsize} \spacingset{1}


\if0\blind
{
  \title{\bf The Ball Pit Algorithm: A Markov Chain Monte Carlo Method Based on Path Integrals}
  \author{ Miguel Fudolig
  \thanks{
    Corresponding Author: mfudolig@huskers.unl.edu
    }
    \hspace{.2cm}\\
    Department of Statistics, University of Nebraska-Lincoln\\
    and \\
    Reka Howard \\
    Department of Statistics, University of Nebraska-Lincoln}
  \maketitle
} \fi

\if1\blind
{
  \bigskip
  \bigskip
  \bigskip
  \begin{center}
    {\LARGE \bf The Ball Pit Algorithm: A Markov Chain Monte Carlo Method Based on Path Integrals}
\end{center}
  \medskip
} \fi

\bigskip
\begin{abstract}
The Ball Pit Algorithm (BPA) is a novel Markov chain Monte Carlo (MCMC) algorithm for sampling marginal posterior distributions developed from the path integral formulation of the Bayesian analysis for Markov chains. The BPA yielded comparable results to the Hamiltonian Monte Carlo as implemented by the adaptive No U-Turn Sampler (NUTS) in sampling posterior distributions for simulated data from Bernoulli and Poisson likelihoods. One major advantage of the BPA is its significantly lower computational time, which was measured to be at least 95\% faster than NUTS in analyzing single parameter models. The BPA was also applied to a multi-parameter Cauchy model using real data of the height differences of cross- and self-fertilized plants. The posterior medians for the location parameter were consistent with other Bayesian sampling methods. Additionally, the posterior median for the logarithm of the scale parameter obtained from the BPA was close to the estimated posterior median calculated using the Laplace normal approximation. The computational time of the BPA implementation of the Cauchy analysis is 55\% faster compared to that for NUTS. Overall, we have found that the BPA is a highly efficient alternative to the Hamiltonian Monte Carlo and other standard MCMC methods.

\end{abstract}

\noindent%
\vfill

\newpage
\spacingset{1.5} 
\section{Introduction}
\label{sec:intro}

One of the key points in performing Bayesian analysis is to sample from the posterior distribution of the model parameters. The posterior distribution, which is the probability density function of the parameter, $\theta$, conditional on the data $X$ can be calculated using the Bayes' rule given by,

\begin{equation}
    P(\theta|X) = \frac{P(X|\theta)P(\theta)}{P(X)},
    \label{eq:bayes}
\end{equation}

where $P(X)$ is the marginal distribution of the data, $P(X|\theta)$ is the likelihood of the data conditional on the parameter, and $P(\theta)$ is the prior distribution of the parameter. The prior distribution P($\theta$) incorporates any prior belief or knowledge about the parameter while the likelihood accounts for the contribution of the data to the value of the parameter \citep{robert2007bayesian}. The posterior distribution is used to make predictions, estimations, and decisions in the Bayesian framework. 

There are special cases where the posterior distribution is described by a standard, well-defined distribution and are thus easy to be sampled from. A good example of these distributions are conjugate priors, which is a group of distributions where the resulting posterior distribution is in the same family of distributions as the defined prior \citep{gelman2013bayesian}. One example of a conjugate prior is the beta prior for a binomial likelihood where evaluating Equation \ref{eq:bayes} when the prior $P(\theta)$ is a beta distribution and the likelihood function $P(X|\theta)$ is a binomial distribution yields a beta distribution for the posterior distribution. An advantage of using conjugate priors is the ease of sampling from standard distributions.

However, not all likelihoods have a conjugate prior distribution and hence, we need to resort to numerical methods for sampling posterior distributions such as Markov chain Monte Carlo (MCMC) methods. MCMC methods use Markov chains of sequential draws of the parameter from approximate distributions to get a good approximation of the target posterior distribution \citep{gelman2013bayesian}. Some well-known MCMC algorithms include the Gibbs sampler \citep{casella1992explaining,geman1984stochastic,smith1993bayesian}, Sequential Monte Carlo (SMC) method \citep{liu1998sequential,doucet2001introduction,del1996nonlinear}, and the Particle Markov Chain Monte Carlo (p-MCMC) method \citep{andrieu2009particle,andrieu2010particle,holenstein2009particle}. The state-space models used in the aforementioned MCMC algorithms simulate a Markov chain in the state space and introduce a criterion for accepting proposal points from the Markov chain based on the target distribution \citep{gelman2013bayesian,yuan2012markov}. The Hamiltonian Monte Carlo (HMC) method \citep{betancourt2017conceptual,duane1987hybrid, hoffman2014no} improves the efficiency of sampling the target distributions by combining MCMC with deterministic simulation methods through incorporating the concept of the Hamiltonian function from physics \citep{gelman2013bayesian}. \cite{girolami2011riemann} improved the efficiency of the HMC by proposing the Riemann Manifold Hamiltonian Monte Carlo (RMHMC), which uses the Riemann geometry of high-dimensional state-spaces to improve the sampling efficiency of the HMC. \cite{lan2015markov} decreased the numerical cost of the RMHMC by proposing an implicit geometric integrator, which resulted to a Lagrangian formulation of the RMHMC.

Another way to decide whether to accept a proposal point is to determine how likely will the chain continue on the trajectory defined by the proposal point. This information can be provided by path integrals, which employs the concept that the total transition probability between two states A and B is the sum of the probabilities of all possible paths from state A to state B \citep{feynman2010quantum,wio2013path,albeverio1976mathematical}. Path integrals are widely used in quantum and statistical physics in investigating time evolution of physical systems \citep{feynman2010quantum,mackenzie2000path,wio2013path,hawking1979path,caldeira1983path,wiegel1975path}, but the concept of path integrals have not been widely utilized in Bayesian analysis. Path integrals were used by \cite{chang2014path} to perform Bayesian inference on physical inverse problems. \cite{chang2015bayesian} used path integrals to develop a non-parametric Bayesian approach to quantify bond energies and mobilities. However, these aforementioned papers did not apply the path integral formulation to Markov chains in state-space models. \cite{fujii2017path} applied the path integral approach to derive analytic expressions of posterior distributions of Markov chains with a specific form of likelihood. However, the derived expressions were not utilized in estimating or making inference on these parameters.

Applying the concept of path integrals to Markov chains in state-space models gives us crucial information about the path taken by the Markov process, which can be used to devise a faster algorithm in sampling posterior distributions. The resulting algorithm is expected to be more efficient compared to the common MCMC methods because the solution of the path integral provides information that predicts the trajectory of the Markov chain in the state-space, which we also refer to as the parameter space, without the need to sample momentum values at each time step.

We developed a new Markov Chain Monte Carlo algorithm to approximate marginal posterior distributions using the resulting equation from the path integral formulation for a Wiener process, a special type of Markov process that describes Brownian motion \citep{szabados_elementary_2010}, in the parameter space. This algorithm was applied on simulated data for Bernoulli and Poisson models and the estimated summary statistics of the posterior distribution was compared to the estimates obtained by the R package \texttt{rstan} \citep{rcite,rstan}. The package \texttt{rstan} implements the Hamiltonian Monte Carlo using an adaptive approach of the No U-Turn Sampler (NUTS) \citep{hoffman2014no}. The resulting algorithm was then applied to a multiparameter Cauchy model with real-world data from the height differences of cross- and self-fertilized plants from the \texttt{R} package \texttt{LearnBayes} \citep{rcite,learnbayes2018}.

Section \ref{sec:path} provides the theoretical basis of the algorithm by introducing the concept of path integrals and explaining its application to the Bayesian analysis of Markov chains. Section \ref{sec:ballpit} explains the proposed algorithm, which we call the Ball Pit Algorithm (BPA), and show how it compares against results from NUTS. This is followed by a discussion of the results after analyzing the simulated data using the BPA and comparing it to the NUTS results.

\section{The Path Integral Formulation}
\label{sec:path}

The path integral method has been used in understanding the evolution of physical systems in time. In quantum mechanics, the path integral can be used in determining how the state of a particle evolves in time by accounting its interactions with its environment \citep{feynman2010quantum,sakurai2011modern}. \cite{wio2013path} worked on the application of path integrals to stochastic processes such as Wiener processes. In this section, we apply the concept of path integrals to Bayesian analysis after briefly introducing what path integrals are.

\subsection{Path Integral}

\begin{figure}[!ht]
    \centering
    \includegraphics[width=3in]{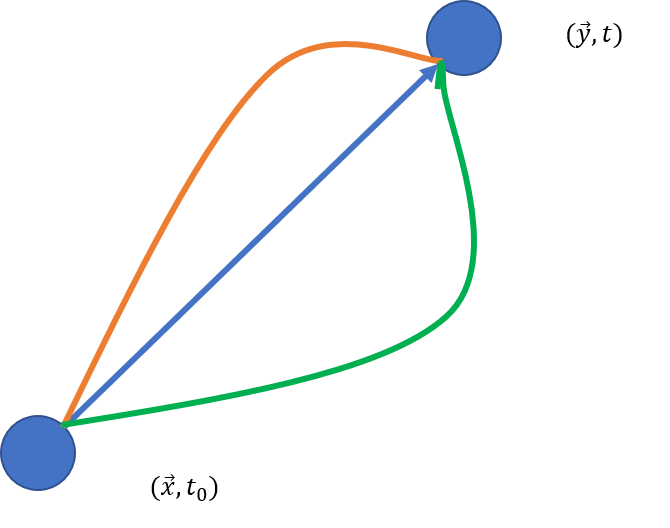}
    \caption{The blue, green, and orange paths are some of the paths that a particle could take from point A to point B from time $t_0$ to $t$.}
    \label{fig:path}
\end{figure}

Consider points A at position $\mathbf{x}$ and B at position $\mathbf{y}$. If one particle starts at point A at time $t_0$ and ends up at point B at time $t$, there are various paths that the particle could have taken from point A to reach point B in between time $t_0$ and $t$. Some of these paths are shown in Figure \ref{fig:path}. The path integral formulation assumes that the total transition probability $P(\mathbf{y},t|\mathbf{x},t_0)$ is the sum of the probabilities of all possible paths the particle could have taken from point A to point B. 

According to \cite{feynman2010quantum}, the conditional probability $P(\mathbf{y},t|\mathbf{x},t_0)$ that the system can have a value of $\mathbf{y}$ at time $t$ given that it had a value of $\mathbf{x}$ at time $t_0$ can be solved by defining a sequence of values $\{x_k\}_{k=1}^N$ such that $x_0 = x$, $x_N = y$, and $t = t_0 + N\epsilon$ where $\epsilon >0$ is the time increment. The resulting formula of the conditional probability is:

    \begin{align}
    P(\mathbf{y},t|\mathbf{x_0},t_0) &= \idotsint dx_1 dx_2 ... dx_{N-1} P(y,t'|x_{N-1},t+\epsilon) \nonumber \\
    &\cdot\cdot\cdot P(x_2,t+2\epsilon|P(x_1,t+\epsilon)P(x_1,t+\epsilon|x_0,t)P(x_0,t).
    \end{align}

This can then be written in terms of the path integral notation as,

\begin{equation}
    P(\mathbf{y},t|\mathbf{x},t_0) = \int \mathcal{D}x[t] \Phi[x(t)]
    \label{eq:pathintdef}
\end{equation}

where 

\begin{equation}
    \mathcal{D}x[t] = \lim_{n\to \infty} dx_1 dx_2 ... dx_{n-1}
\end{equation}

and 
\begin{equation}
    \Phi[x(t)] = \lim_{\epsilon \to 0} \prod_{i=1}^N P(x_i, t+i\epsilon|x_{i-1}, t+(i-1)\epsilon)
\end{equation}

The term $\Phi[x(t)]$ is referred to as the propagator \citep{sakurai2011modern}, which can be interpreted as a probability of a specific path the system follows from $x_0$ to $x_N$. For stochastic processes such as Brownian motion and Markov chains, Equation \ref{eq:pathintdef} can be simplified into the following form,

\begin{equation}
    P(\mathbf{y},t|\mathbf{x},t_0) = \int \mathcal{D}x[t] \exp[-S_E]
    \label{eq:pathint}
\end{equation}

where $S_E$ is the discrete-time Euclidean action can be defined as the following,

\begin{equation}
    S_E = \sum_{i=1}^N \mathcal{L}(x_i,\dot{x}_i,t_i)\Delta t_i
\end{equation}

where the integration runs over time from $t=t_1$ to $t'=t_N$ and $\dot{x} = \dfrac{x(t) - x(t-\epsilon)}{\epsilon}$ is the speed of the system. The quantity $\mathcal{L}$ is defined as the Lagrangian of the system, which is a key function in physics in describing the motion of the system \citep{goldstein2002classical}. The Lagrangian of the system is expressed as the difference between the kinetic energy $K$ and the potential energy $U$ \citep{goldstein2002classical}. The kinetic energy is associated with the energy in motion while the potential energy is associated with the configuration of the system \citep{tipler2007physics}. Once the Lagrangian function of a process is established, the trajectory of the process is well-defined given that its initial position and velocity is known. 

We can apply these concepts to Bayesian analysis by utilizing path integrals to describe the trajectory of Markov chains used in MCMC methods. The path integral formulation of Bayesian analysis of Markov chains is explained further in the next subsection.

\subsection{Path Integrals: Bayesian Analysis of Markov Chains}
\label{sec:bayes}

\cite{fujii2017path} used the path integral approach for Markov chains in Bayesian analysis. One of the limitations of their study was that their approach only applied to likelihoods with the form:

\begin{equation}
    L(\theta|x) = M \exp\left[-V(\theta - x(t))\epsilon\right],
\end{equation}

where $\theta$ is the parameter, $x(t)$ is the data, $\epsilon$ is the time increment between the steps of the Markov chain, $V$ is a function of the difference of the parameter and the data, and $M$ is the normalization constant. This form of the likelihood is similar to exponential family likelihood except for the explicit dependence on the time increment $\epsilon$ and the difference between parameter and data, which are not commonly encountered in statistical models. We derive the path integral formulation for a more general form of the likelihood and study the implications of the resulting expression.

We start with the concept of recursive Bayesian estimation for Markov chains outlined by \cite{chen2003bayesian}. Suppose we have a parameter $\theta_N$ at time step $N$ that follows a first-order Markov process, i.e. $\theta_N$ is only conditional on the value of the immediately preceding time step and independent of the rest of the steps in the path. We denote $\mathbf{X}_N = (x_0,x_1,...,x_N)$ as the vector of independent observations from time-step $t=0$ to $t = N$. Then the posterior distribution of $\theta_N$ conditional on the observations $\mathbf{X}_N$ can be solved using the Bayes' rule.

\begin{eqnarray}
    P(\theta_N|\mathbf{X}_N) &=&\dfrac{P(\mathbf{X}_N|\theta_{N})P(\theta_N)}{P(\mathbf{X}_N)}\\
    &=& \dfrac{P(x_N,\mathbf{X}_{N-1}|\theta_{N})P(\theta_N)}{P(x_N,\mathbf{X}_{N-1})}\\
    &=& \dfrac{P(x_N|\theta_N,\mathbf{X}_{N-1})P(\mathbf{X}_{N-1}|\theta_N)P(\theta_N)}{P(x_N|\mathbf{X}_{N-1})P(\mathbf{X}_{N-1})}\\
    &=& \dfrac{P(x_N|\theta_N)P(\theta_N|\mathbf{X}_{N-1})}{P(x_N|\mathbf{X}_{N-1})} \label{eq:reqbe}
\end{eqnarray}

The term $P(\theta_N|\mathbf{X}_{N-1})$ is the probability density of the parameter conditioned on the observations before the current time step, which implies that this term is a conditional prior for $\theta_N$. This probability can be expressed as the following integral:

\begin{equation}
    P(\theta_N|\mathbf{X}_{N-1}) = \int P(\theta_N|\theta_{N-1})P(\theta_{N-1}|X_{N-1})d\theta_{N-1},
    \label{eq:recprior}
\end{equation}

Substituting \ref{eq:recprior} into \ref{eq:reqbe},

\begin{equation}
    P(\theta_N|\mathbf{X}_N)= \dfrac{P(x_N|\theta_N)}{P(x_N|\mathbf{X}_{N-1})} \int P(\theta_N|\theta_{N-1})P(\theta_{N-1}|\mathbf{X}_{N-1})d\theta_{N-1}
    \label{eq:start}
\end{equation}

where the integral is on the support of $\theta_{N-1}$. The term $P(\theta_N|\theta_{N-1})$ can be interpreted as the transition probability of the parameter from one time step to another. We consider the parameter $\theta$ as a random variable that evolves as described by the Brownian motion stochastic differential equation given by:

\begin{equation}
    d\theta = \mu(\theta,t) dt + \sigma dW_t,
    \label{eq:sde}
\end{equation}

where $\mu(\theta,t)$ is the drift velocity of the stochastic process and $\sigma$ is the strength of the noise. $W_t$ is known as the Wiener process at time $t$, which is the continuous model of Brownian motion. The Wiener process can also be defined in terms of the following distribution \citep{szabados_elementary_2010, wio2013path},

\begin{equation}
    W_t - W_s \sim N(0,s),
\end{equation}

where $s$ is a time point such that $0 \leq s < t$. When the drift velocity is set to zero, Equation \ref{eq:sde} reduced to a Wiener process for $\theta$. Based on the definition of the Wiener process, the distribution of $\theta_{t+dt}$ for any time change $dt > 0$,

\begin{equation}
    \theta_{t+dt}|\theta_{t}\sim N(\theta_t,\sigma^2 dt) 
\end{equation}

This can also be written in equation form,

\begin{equation}
    P(\theta_{t+dt}|\theta_{t}) = (2\pi\sigma^2 dt )^{-1/2} e^{-\frac{\dot{\theta}_{t+dt}^2}{2\sigma^2}dt} 
    \label{eq:discnodrift}
\end{equation}

where $\displaystyle \dot{\theta}_{t+dt} = \frac{\theta_{t+dt} - \theta_t}{dt}$. 







We consider $dt=1$, which corresponds to the discrete case of the Markov process. Substituting Equation \ref{eq:discnodrift} to Equation \ref{eq:start} yields the following integral:

\begin{equation}
      P(\theta_N|\mathbf{X}_N)= \dfrac{P(x_N|\theta_N)}{P(x_N|\mathbf{X}_{N-1})(2\pi\sigma^2 )^{1/2}} \int{ e^{-\dfrac{(\theta_{t+1}-\theta_t)^2}{2\sigma^2}} P(\theta_{N-1}|\mathbf{X}_{N-1})d\theta_{N-1}}  
\end{equation}

This can also be written as,

\begin{equation}
      P(\theta_N|\mathbf{X}_N)= C \int{ e^{\left[-\dfrac{(\theta_{N}-\theta_{N-1})^2}{2\sigma^2} + \log P(x_N|\theta_N)\right]} P(\theta_{N-1}|\mathbf{X}_{N-1})d\theta_{N-1}},
      \label{eq:recursionnd}
\end{equation}

where all the terms constant with respect to $\theta_N$ are pooled into a general normalizing constant $C$. Equation \ref{eq:recursionnd} shows a recursive pattern between the posterior distributions at time steps $N-1$ and $N$. Continuing this recursion, we get the following integral,

\begin{equation}
      P(\theta_N|\mathbf{X}_N,\theta_0)= C \idotsint \prod_{i=0}^{N-1} d\theta_{i} e^{\left[-\dfrac{(\theta_{i+1}-\theta_i)^2}{2\sigma^2} + \log P(x_{i+1}|\theta_{i+1})\right]} P(\theta_{0}),
      \label{eq:recursionnd2}
\end{equation}

We focus on the product in the integrand which can be written as,

\begin{equation}
      \prod_{i=0}^{N-1} d\theta_{i} e^{\left[-\dfrac{(\theta_{i+1}-\theta_i)^2}{2\sigma^2} + \log P(x_{i+1}|\theta_{i+1})\right]}  = \exp\left[-\sum _{i=1}^N \left(\dfrac{\dot{\theta}_i^2}{2\sigma^2} - \log P(x_{i}|\theta_{i})\right)\right]
      \label{eq:pbref}
\end{equation}



The above equation is similar to the definition of the Wiener path integral related to the Wiener measure \citep{glimm2012quantum,wio2013path}. Comparing the terms to Equation \ref{eq:pathint}, we determine the discrete Euclidean action for the parameter space of $\theta$ is given by,

\begin{equation}
    S_E = \sum _{i=1}^N \left(\dfrac{\dot{\theta}_i^2}{2\sigma^2} - \log P(x_{i}|\theta_{i})\right)
    \label{eq:action}
\end{equation}




Based on Equation \ref{eq:action} and the idea that the process should accelerate towards the point of maximum likelihood, the corresponding Lagrangian to determine the equation of motion of the process is given by,

\begin{equation}
  \mathcal{L} = \left[\dfrac{\dot{\theta}_\tau^2}{2\sigma^2} + \left(\log P(x_\tau|\theta_\tau)\right)\right]
  \label{eq:lagrangian}
\end{equation}

 Classical physics tells us that the process will evolve by taking the path of least action. Since the Euclidean action is dependent on the likelihood, this implies that points simulated from the prior will evolve in the parameter space as dictated by the value of the log-likelihood. As we reach equilibrium, the prior distribution is expected to have evolved to the posterior distribution as guided by the likelihood.

The first term of Equation \ref{eq:lagrangian} can be interpreted as the energy due to the inherent motion of the process due to its stochastic nature. A process with a high  $\sigma$ value will have a larger variance in its trajectory and hence will cover a wider area in the parameter space. This is analogous to the smaller masses accelerating faster when a force is applied to them, which is consistent with the interpretation of noise strength by \cite{lan2015markov} in their work on Lagrangian Monte Carlo. 

The log-likelihood in Equation \ref{eq:action} incorporates the information from the observations on choosing the path taken by the process. We expect the process to accelerate towards the value of maximum likelihood or equivalently, the value of the minimum negative log-likelihood. This implies that the process should be pulled towards this point at any given time. Using the negative log-likelihood as the counterpart for the potential energy instead of the positive makes the dynamics of the process consistent with what is expected in physical systems. We use this idea in formulating the equations of motion in the parameter space.

\subsection{Equations of Motion}

The Lagrangian can be used to find the equations of motion of the system \citep{goldstein2002classical} which, in the context of this research, will describe how the system moves in the parameter space upon recording an observation, $x$, at a given time $\tau$. Assuming that our likelihood is explicitly independent of the speed of the process, $\dot{\theta}$, and $\tau$, the equations of motion for the parameter is given by the Euler-Lagrange equation \citep{goldstein2002classical},

\begin{equation}
    \dfrac{d}{dt}\left(\dfrac{\partial \mathcal{L}}{\partial \dot{\theta}}\right) - \dfrac{\partial \mathcal{L}}{\partial \theta} = 0
    \label{eq:eom}
\end{equation}

This principle is similar to the Hamiltonian equations of motion that we use in Hamiltonian Monte Carlo \citep{betancourt2017conceptual}, but instead of sampling momentum values for every time step, the Euler-Lagrange equation can be used to update the speed of the process in the state space. This result describes how the process moves in the $\theta$-space, which can be interpreted as the parameter space in the context of Bayesian analysis. The form of Equation \ref{eq:lagrangian} is similar to the Lagrangian function used by \cite{lan2015markov} in their work on Explicit Riemannian Manifold Lagrangian Monte Carlo. The resulting equation is a second order ordinary differential equation (ODE) in time which is easily solvable using numerical methods.

We use this form for the equation of motion of the process to create an algorithm that samples from a target posterior distribution by evolving points sampled from the prior in an MCMC framework.

\section{The Ball Pit Algorithm (BPA): An Approximate Bayesian Method to Estimate Posterior Statistics}
\label{sec:ballpit}

In this section, we discuss how a Monte Carlo algorithm to approximate posterior distributions was developed using the Lagrangian function and the equation of motion we derived in Section \ref{sec:bayes}.

\subsection{Algorithm}
The main idea of the algorithm is that the prior distribution evolves to the posterior distribution by taking the path dictated by the value of the likelihood. A Wiener process moving in the parameter space will follow the trajectory dictated by the equation of motion derived from the Lagrangian function. 

A pseudocode of the algorithm is presented in Algorithm 1. The algorithm starts by simulating starting points from the prior distribution, an approach used by \cite{pompe2020framework} in their adaptive MCMC algorithm for multimodal distributions. These points will be utilized as starting points for a single independent process which we will refer to as a "ball". Each ball will be given an initial speed, $\dot{\theta}$, which will be simulated from a Gaussian distribution with centered at zero and variance equal to the noise strength of the Markov process $\sigma^2$. The trajectory of each ball, which we refer to as the chain, in the state space will be recorded by numerically solving the differential equation in Equation \ref{eq:eom}. In the evolution of these chains, we need to avoid moving on paths unlikely for the chain to follow. Hence, we need to introduce a stopping criterion such that when the criterion is satisfied, the ball would change its trajectory to avoid following an unlikely path. This stopping criterion will be based on the probability of following the proposed path, $E$, given by,

\begin{equation}
    E(\theta,\dot{\theta})= \exp\left[{-\epsilon  \frac{\dot{\theta}^2}{2\sigma^2} + \log L(X|\theta)}\right],
\end{equation}

where $\epsilon$ is the time increment between steps in solving the equation of motion and $\log L(X|\theta)$ is the log-likelihood. In the context of the path integral formulation, the effective weight of the proposed path is the propagator. We impose the stopping criterion for a candidate point $(\theta^*,\dot{\theta}^*)$ given a current point $(\theta_0,\dot{\theta}_0)$ in a form of an acceptance probability, $p_{accept}$, defined as:

\begin{equation}
    p_{accept} = \min\left(1,\frac{E(\theta^*,\dot{\theta^*})}{E(\theta_0,\dot{\theta_0})}\right)
\end{equation}

The term $\displaystyle \frac{E(\theta^*,\dot{\theta^*})}{E(\theta_0,\dot{\theta_0})}$ can be interpreted as a risk ratio in continuing the movement along the proposed path. The candidate point will be accepted with probability $p_{accept}$. If the candidate point is rejected, the process will remain at $\theta_0$ and its speed will be re-sampled from the same Gaussian distribution where the initial speeds were sampled from. 

There are situations where the value for $\theta^*$ is rejected for each step depending on the choice for the prior distribution. To avoid the ball being stuck at an unlikely parameter value, we perform a check for every rejection if the point has been rejected for a user-defined lag time $\tau$. If this is the case, the ball will be given a new starting position by sampling from the prior distribution again. 

All balls are allowed to move until a final evolution time $T$ when the desired chain length is reached. After accounting for a burn-in phase for each chain, the chains are combined to form an approximate posterior distribution. We call this algorithm the Ball Pit Algorithm (BPA) because the posterior distribution is approximated by the distribution of the balls, which are assumed to cover the relevant area of the parameter space.

This algorithm was tested on single parameter models and the posterior statistics obtained from the BPA will be compared to the results of the adaptive No U-Turn Sampler (NUTS) algorithm, which is commonly used to implement HMC. The BPA was also used to analyze real-world data of plant heights of self- and cross-fertilized plants by sampling marginal posterior distributions for a Cauchy multi-parameter model.

\begin{algorithm}
\caption{The Ball Pit Algorithm: One Process}\label{euclid}
\begin{algorithmic}[1]
\Procedure{Path Integrals}{}
\State $\text{Initialize}~\textit{prior,}~\sigma^2,\epsilon,t=0,T,\tau$
\State $\theta_0 \sim \textit{prior}$
\State $\dot{\theta_0} \sim N(0,\sigma^2)$
\BState \emph{Chain Start}:
\State $\theta \gets \theta_0$
\State $v \gets \dot{\theta_0} $
\BState \emph{Update}:
\State $\text{Update}~(\theta,v)~\text{using Euler-Lagrange Equations to get}~(\theta^*,v^*) $
\If {$L(\theta^*)~\text{is NA}$}  
\State $\text{keep } \theta $
\State $v \sim N(0,\sigma^2)$
\EndIf
\State $p_{acc} \gets \min(1,\exp{\left[E(\theta^*,v^*)-E(\theta,v)\right]})$
\State $p \sim Unif(0,1)$
\If {$p < p_{acc}$} 
\State $\theta \gets \theta^*$
\State $v \gets v^* $
\Else {
 \If {$\theta = \theta_{(t-\tau)}$}
\State $\text{Resample } \theta \text{ from prior.}$
\State $v \sim N(0,\sigma^2)$
 \Else{
\State $\text{keep}~\theta$
\State $v \sim N(0,\sigma^2)$}
\EndIf
}
\EndIf
\State $t \gets t+\epsilon $
\If {t = T}
\State \textbf{End Process}
\Else {\State \textbf{go to} \emph{Update}}.
\EndIf
\EndProcedure
\end{algorithmic}
\end{algorithm}

\subsection{Advantages}
\label{subsec:adv}
The proposed method is similar to the Hamiltonian Monte Carlo, which is also a method derived from physics concepts. The Hamiltonian Monte Carlo (HMC) involves simulating "momentum" values, $p$, at the start of each iteration from a conditional normal distribution to determine the gradient of steepest descent of the target distribution in the phase space ($(\theta,p)$ space). The proposed value of the parameter is obtained from the Hamiltonian equation of motion \citep{betancourt2017conceptual}. The integration of physics in the HMC decreases the noise due to the inherent stochasticity of the Markov chain, making it more efficient compared to the other MCMC methods \citep{gelman2013bayesian}.

In contrast, the BPA follows a Lagrangian framework where the position and the velocity of the process evolves as dictated by the Euler-Lagrange equation (Equation \ref{eq:eom}). New velocities for each ball need to be sampled only when the chain starts or the proposed point is rejected, which means there are fewer sampling steps in the BPA compared to the HMC. This translates to faster convergence of the algorithm and lower computational time compared to the HMC.

\section{Bayesian Analysis Using the BPA: Simulated and Real-World Data}

This section discusses the results obtained from using the Ball Pit Algorithm (BPA) to approximate the posterior distribution of single parameter models. In this research, we considered single parameter models with Bernoulli and Poisson likelihoods. The Bernoulli likelihood was chosen because of its parameter has a bounded support and the existence of a conjugate prior. The Jeffreys prior for the Poisson distribution is an improper prior, which enabled us to explore how improper priors can affect the resulting approximation for the posterior distributions. Posterior statistics and the computation time for the algorithm were recorded and then compared to the results obtained through the NUTS algorithm included in the R package \texttt{rstan} \citep{rstan,rcite}. 

The BPA was implemented using 80 balls that are made to move around the state space for an evolution time of 1 second with a time step size of 0.01 seconds to match the default overall chain length used by \texttt{rstan} \citep{rstan}. The values for the first five seconds, which translates to a chain length of 500 for each ball, will be treated as a warm-up phase and will be discarded just as in the regular MCMC algorithm. The sampled marginal posterior distribution was then compared to a NUTS implementation that used 80 chains with a chain length of 100.

We then tested the BPA to sample marginal posterior distributions from a multi-parameter model using a real-world data set from the \texttt{LearnBayes} package used by \cite{albert_bayesian_2009} that contains the height differences between cross- and self-fertilized plants.

\subsection{Bernoulli Process}

The corresponding likelihood for a sample size $N$ for a Bernoulli process with probability $\theta$ can be written as,

\begin{equation}
    P(X|\theta) = \prod_{i=1}^N \theta^{x_i}(1-\theta)^{N-x_i}
\end{equation}

We used Equation \ref{eq:lagrangian} to derive the Lagrangian function with the given likelihood function with probability $\theta$, size $N$ and observation $\mathbf{x} = (x_1,x_2,...x_N)$.

\begin{equation}
       \mathcal{L} = \left[\frac{\dot{\theta }^2}{2\sigma^2} + \left(\sum_{i=1}^N x_i \log \theta + \left(N-\sum_{i=1}^N x_i\right) \log(1-\theta)\right)\right]
\end{equation}

Then the corresponding equation of motion for $\theta$ can be solved from Equation \ref{eq:eom}:

\begin{eqnarray}
    \dfrac{d}{dt}\left(\dfrac{\dot{\theta}}{\sigma^2}\right) - \left[ \dfrac{\sum_{i=1}^N x_i}{\theta} - \dfrac{\left(N-\sum_{i=1}^N x_i\right)}{1-\theta}\right] &=& 0 \\
    \ddot{\theta} - \sigma^2 \left[\dfrac{\sum_{i=1}^N x_i-N\theta}{\theta(1-\theta)}\right] &=& 0 \label{eq:eombin}
\end{eqnarray}

where $\theta \in (0,1)$. $\dot{\theta}$ and $\ddot{\theta}$ are the velocity and acceleration of the process, respectively. Solving the second-order differential equation in Equation \ref{eq:eombin} would provide the candidate values for the parameter $\theta$ and velocity $\dot{\theta}$ of the process.

\begin{figure}[!ht]
    \centering
    \includegraphics[width=3in]{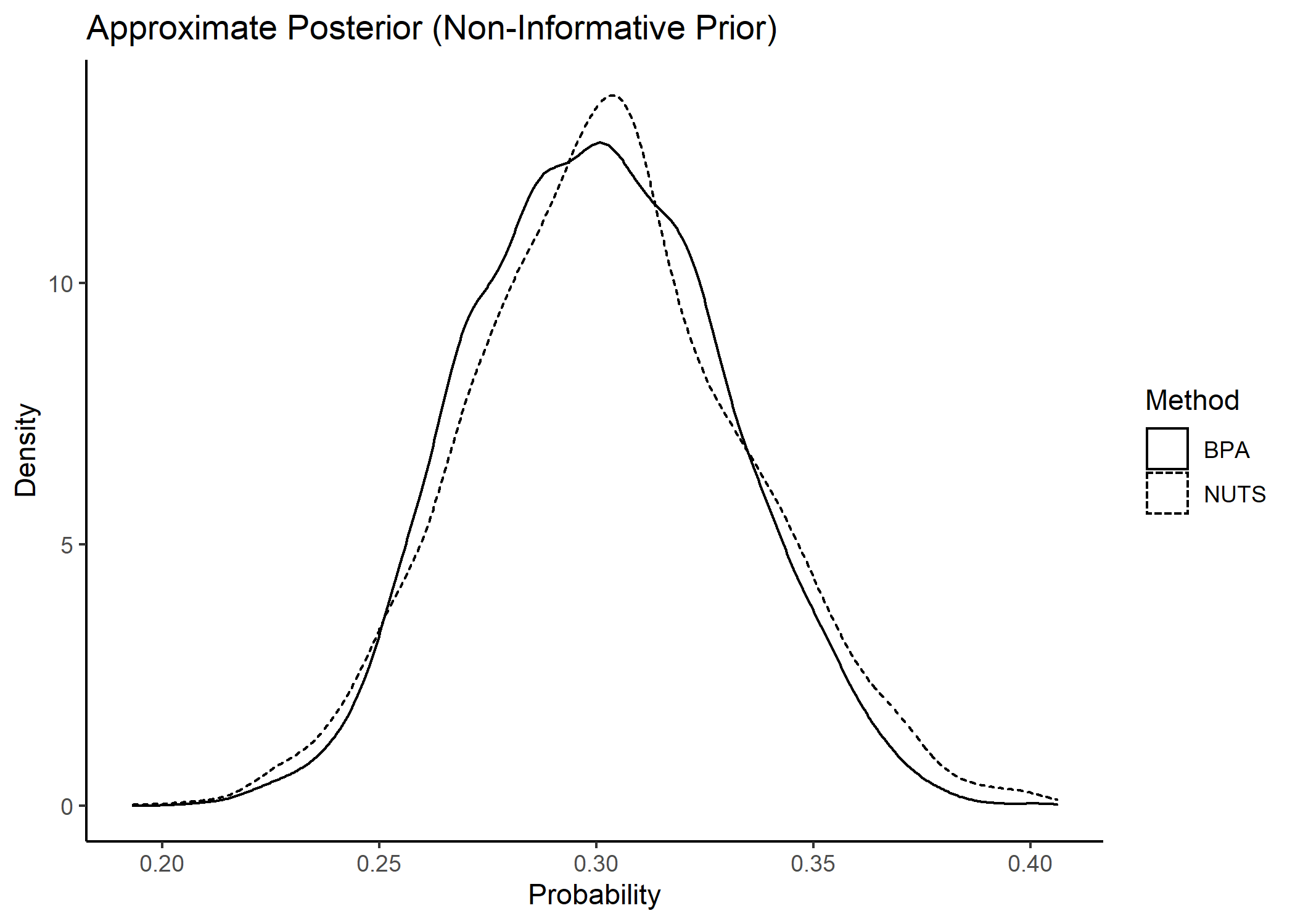}
    \caption{The samples from the posterior distribution from the BPA and NUTS method for the uniform prior. The total number of post warm-up samples for both methods is 4000.}
    \label{fig:bin_unif}
\end{figure}

\begin{figure}[!ht]
    \centering
    \includegraphics[width=3in]{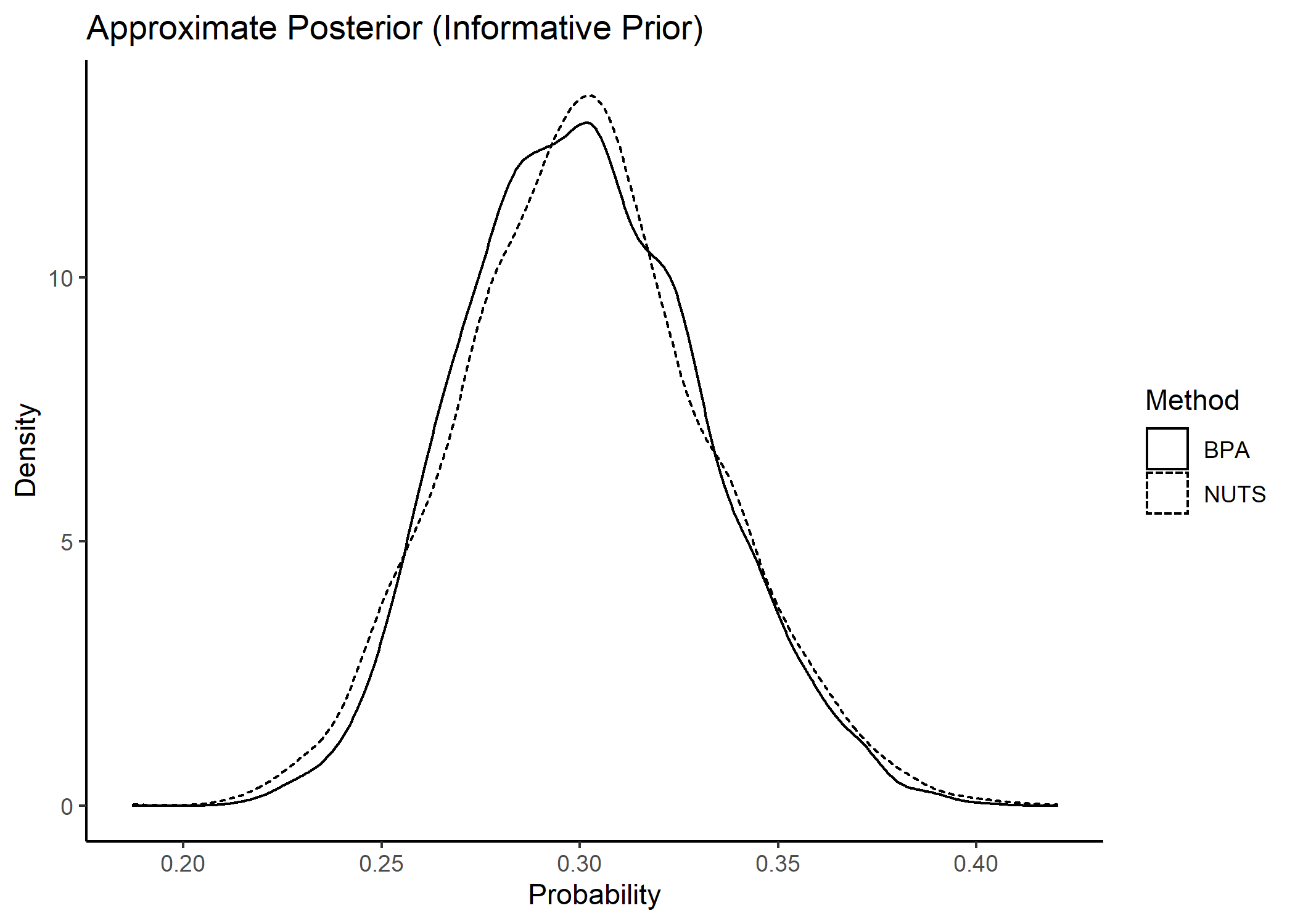}
    \caption{The samples from the posterior distribution from the BPA and NUTS method for an informative Beta(3,7) prior. The total number of post warm-up samples for both methods is 4000.}
    \label{fig:bin_inf}
\end{figure}

For the simulation study, a representative data set of size 200 was simulated from a Bernoulli distribution with probability set to a representative value $\theta=0.3$. We set the Markov process variance to be 1. Two types of priors will be considered for this analysis to see whether the consistency of the BPA compared to the NUTS algorithm depends on the choice of prior. Figures \ref{fig:bin_unif} and \ref{fig:bin_inf} and Tables \ref{tab:bin_inf} show the respective approximate posterior distributions and quantiles for the non-informative and informative prior for $\theta$. The non-informative prior used is $Beta(1,1)$, which is the same as the $Unif(0,1)$ distribution, while the informative prior used is $Beta(3,7)$. The beta distribution was selected to be the choice of prior because it is a conjugate prior to a Bernoulli likelihood.



\begin{table}[H]
\centering
\caption{Estimates of the posterior distribution from theoretical Bayes,  NUTS, and the Ball Pit Algorithm (BPA) with time elapsed for a uniform prior ($Beta(1,1)$).} 
\label{tab:bin_unif}
\begin{adjustbox}{width=1\textwidth}
\begin{tabular}{|c|c|c|c|c|c|c|c|c|c|c|}
\hline
\textbf{Method}                         & Mean & Std. Error.     & Std. Dev. & 2.5 \% & 25 \% & 50 \% & 75 \% & 97.5 \% & Time (s)   \\ \hline
\textbf{NUTS}  & 0.30 & \textless{}0.01 & 0.03      & 0.24   & 0.28  & 0.3  & 0.32 & 0.37    & 34.23 \\ \hline
\textbf{BPA } & 0.30 & \textless{}0.01    & 0.03      & 0.25   & 0.28  & 0.3  & 0.32  & 0.36    & 0.71 \\ \hline
\end{tabular}
\end{adjustbox}

\end{table}



\begin{table}[H]
\centering
\caption{Estimates of the posterior distribution from NUTS and the Ball Pit Algorithm (BPA) with time elapsed for an informative prior Beta(3,7). } 
\label{tab:bin_inf}
\begin{adjustbox}{width=1\textwidth}
\begin{tabular}{|c|c|c|c|c|c|c|c|c|c|c|}
\hline
\textbf{Method}                         & Mean & Std. Error.     & Std. Dev. & 2.5 \% & 25 \% & 50 \% & 75 \% & 97.5 \% & Time  \\ \hline
\textbf{NUTS}                           & 0.3  & \textless{}0.01 & 0.03      & 0.24   & 0.28  & 0.30  & 0.32  & 0.37    & 29.39 \\ \hline
\textbf{BPA} & 0.3 & \textless{}0.01 & 0.03     & 0.25   & 0.28  & 0.3  & 0.32  & 0.37    & 0.63  \\ \hline
\end{tabular}
\end{adjustbox}
\end{table}

The two methods yielded similar approximations of the posterior distribution as displayed in the plots in Figures \ref{fig:bin_unif} and \ref{fig:bin_inf}. The quantile estimates of the BPA are also close to the results provided by NUTS. The big difference between the two methods lie in the computational time. The computational time was reduced by 98\% in generating the approximate posterior distribution for the uniform prior, while the computational time was reduced by 97\% for the informative prior.

\subsection{Poisson Process}

If the response follows a Poisson distribution with a constant rate parameter $\lambda$, then the likelihood can be written  as

\begin{equation}
    P(X|\theta) = \prod_{i=1}^N e^{-\lambda} \frac{\lambda^{x_i}}{x_i!}.
\end{equation}

The corresponding Lagrangian is given by,

\begin{equation}
    \mathcal{L} = \left[\dfrac{\dot{\lambda }^2}{2\sigma^2} + \left(-\lambda + x \log \lambda - \log \left(x!\right)\right)\right]
\end{equation}

The equation of motion based on this Lagrangian can be calculated by taking the logarithm of this likelihood and substituting to Equation \ref{eq:lagrangian},

\begin{eqnarray}
    \dfrac{d}{dt}\left(\dfrac{\dot{\lambda}}{\sigma^2}\right) - \left[ \dfrac{x}{\lambda} - 1\right] &=& 0 \\
    \ddot{\lambda} - \sigma^2 \left[\dfrac{x-\lambda}{\lambda}\right] &=& 0 \label{eq:eompois}
\end{eqnarray}

Similar to Equation \ref{eq:eombin}, the solution of Equation \ref{eq:eompois} is well-defined if the initial values of the rate parameter value $\lambda$ and velocity $\dot{\lambda}$ are set.

\begin{figure}[!ht]
    \centering
    \includegraphics[width=3in]{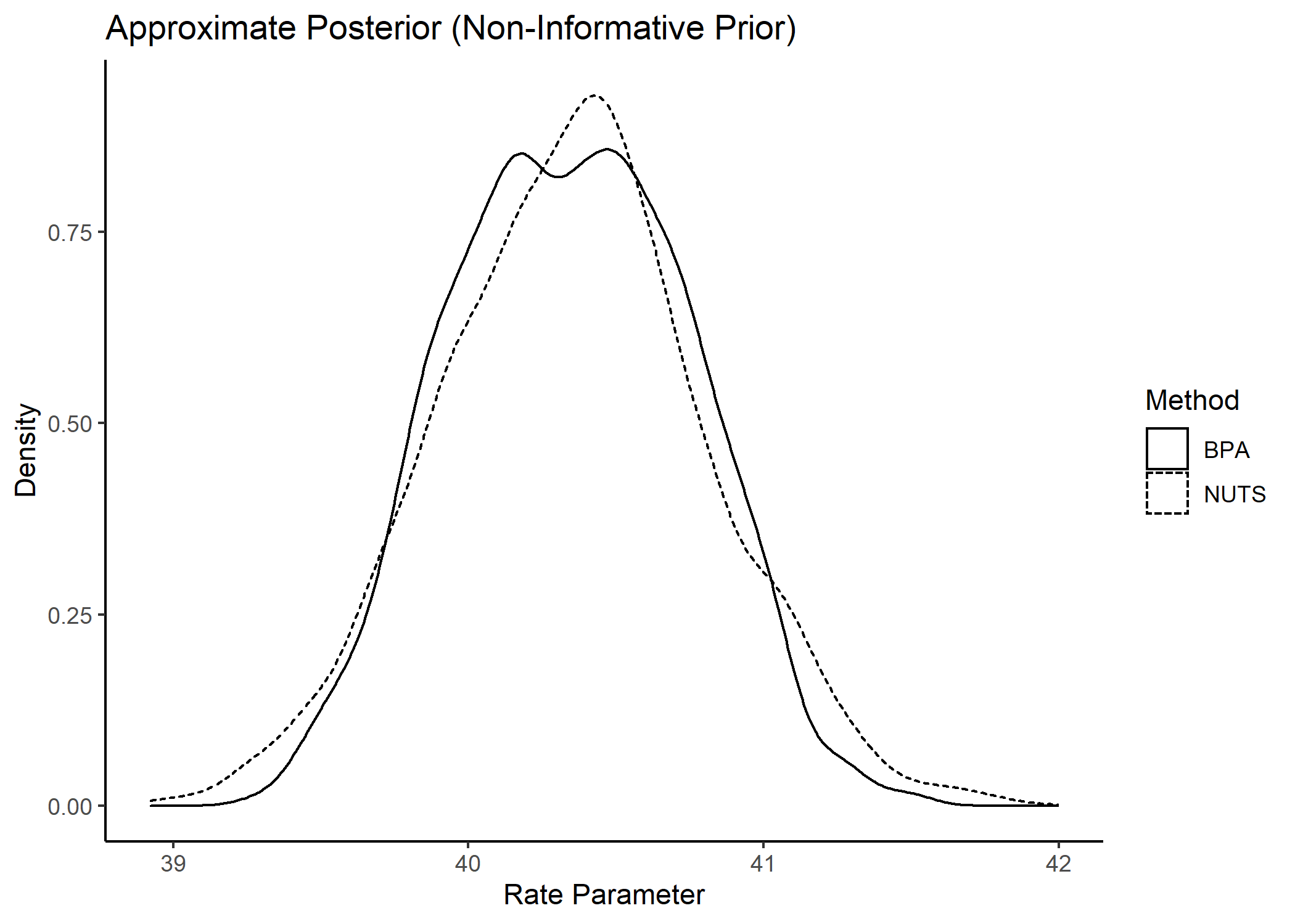}
    \caption{The samples from the posterior distribution from the BPA and NUTS method for the Jeffreys prior. The total number of post warm-up samples for both methods is 4000.}
    \label{fig:poisson_uninf}
\end{figure}

\begin{figure}[!ht]
    \centering
    \includegraphics[width=3in]{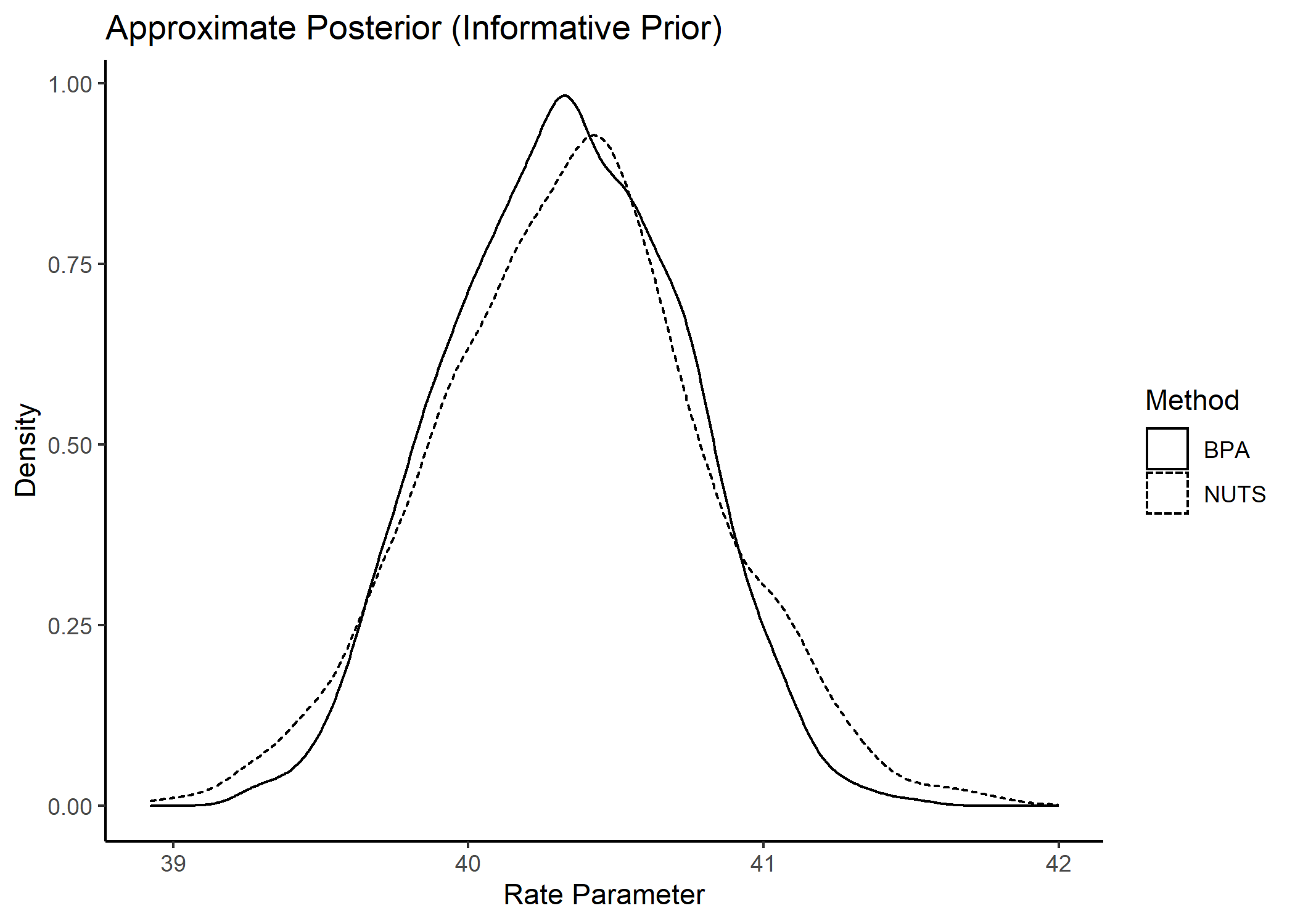}
    \caption{The samples from the posterior distribution from the BPA and NUTS method for an informative $N(\bar{x},4)$ prior. The total number of post warm-up samples for both methods is 4000.}
    \label{fig:poisson_inf}
\end{figure}

For the simulation study, a data set of size 200 was simulated from a Poisson distribution with a rate parameter $\lambda = 40$. We used Equation \ref{eq:eompois} to update the position and the speed of the process in the BPA. The value of the Markov process variance was set to $\sigma^2 = 100$. We chose a larger value for $\sigma^2$ because we want the process to have less "mass" to cover a wider area of the parameter space for the rate parameter, which is larger compared to the parameter space of the Bernoulli probability in the previous example. Similar to the approach we used for the Bernoulli likelihood example, we now compare the performance of the BPA compared with NUTS for non-informative and informative priors on a simulated Poisson data set. The Jeffreys prior was used as the non-informative prior for the Poisson rate parameter, $\lambda$, which is given by $P(\lambda) \propto \lambda^{-1/2}$. Meanwhile, the informative prior used was a normal distribution centered at the sample mean with a variance of 4. The density plots of the approximate distributions for both priors are shown in Figures \ref{fig:poisson_uninf} and \ref{fig:poisson_inf}, and the quantile estimates are listed in Tables \ref{tab:pois_jeff} and \ref{tab:pois_inf}.

\begin{table}[H]
\centering

\caption{Estimates of the posterior distribution from theoretical Bayes,  NUTS, and the Ball Pit Algorithm (BPA) with time elapsed for the Jeffreys prior $P(\lambda) \propto \lambda^{-1/2}$. } 
\begin{adjustbox}{width=1\textwidth}
\begin{tabular}{|c|c|c|c|c|c|c|c|c|c|c|}
\hline
\textbf{Method}                          & Mean  & Std. Error. & Std. Dev. & 2.5 \% & 25 \% & 50 \% & 75 \% & 97.5 \% & Time  \\ \hline
\textbf{NUTS}                            & 40.35 & 0.01        & 0.46      & 39.45  & 40.04 & 40.36 & 40.64 & 41.26   & 37.41\\ \hline
\textbf{BPA} & 40.34 & 0.08       & 0.40      & 39.58  & 40.04 & 40.34 & 40.64 & 41.08   & 1.17 \\ \hline
\end{tabular}
\label{tab:pois_jeff}
\end{adjustbox}
\end{table}

\begin{table}[H]
\centering
\caption{Estimates of the posterior distribution from NUTS and the Ball Pit Algorithm (BPA) with time elapsed for an informative $N(\bar{x},4)$ prior.} 
\begin{adjustbox}{width=1\textwidth}
\begin{tabular}{|c|c|c|c|c|c|c|c|c|c|c|}
\hline
\textbf{Method}                          & Mean  & Std. Error. & Std. Dev. & 2.5 \% & 25 \% & 50 \% & 75 \% & 97.5 \% & Time  \\ \hline
\textbf{NUTS}                            & 40.36 & 0.01        & 0.44      & 39.53  & 40.07 & 40.35 & 40.65 & 41.23   & 33.22 \\ \hline
\textbf{BPA} & 40.32 & 0.09        & 0.39      & 39.60  & 40.04 & 40.33 & 40.61 & 41.05   & 1.06 \\ \hline
\end{tabular}
\label{tab:pois_inf}
\end{adjustbox}
\end{table}

  The density plots for the posterior distribution show that the approximated posterior distribution obtained using the BPA is narrower compared to the NUTS result. This observation was supported by the smaller standard deviation and narrower credible intervals for the BPA chain in Tables \ref{tab:pois_jeff} and \ref{tab:pois_inf}. The difference in width between the two approximations was more pronounced for the informative prior, which might be explained by the mechanism of the algorithm. Since the starting points are sampled from the prior, the balls are localized in the neighborhood of the minimum point of the negative log-likelihood. This made it unlikely for the chains to explore points far from the mean, resulting to a narrower distribution from the sampled points. Similar to the Bernoulli likelihood example, the computational time was significantly reduced by 97\% for both priors.

\subsection{Multiparameter Model: Cauchy Distributed Errors}

We tested the BPA to sample marginal posterior distributions from a multi-parameter model. We considered a real-world data set from the \texttt{LearnBayes} \citep{learnbayes2018} package used by \cite{albert_bayesian_2009} that contains the height differences between cross- and self-fertilized plants. These differences are modeled by a Cauchy likelihood with a location parameter $\mu$ and scale parameter $\sigma$. 

The approach we took in applying the BPA to a multiparameter model is similar to Gibbs sampling where we make an educated guess about the parameter values and sequentially updating these values through implementing the algorithm recursively. We used the \texttt{laplace} function in the \texttt{LearnBayes} \citep{learnbayes2018} package to get the an estimate of the posterior modes for both the location parameter and the logarithm of the scale parameter. We then use these values as an input to the BPA as it simulates the marginal posterior distribution for both parameters. 

\begin{figure}[!ht]
    \centering
    \includegraphics[width=3in]{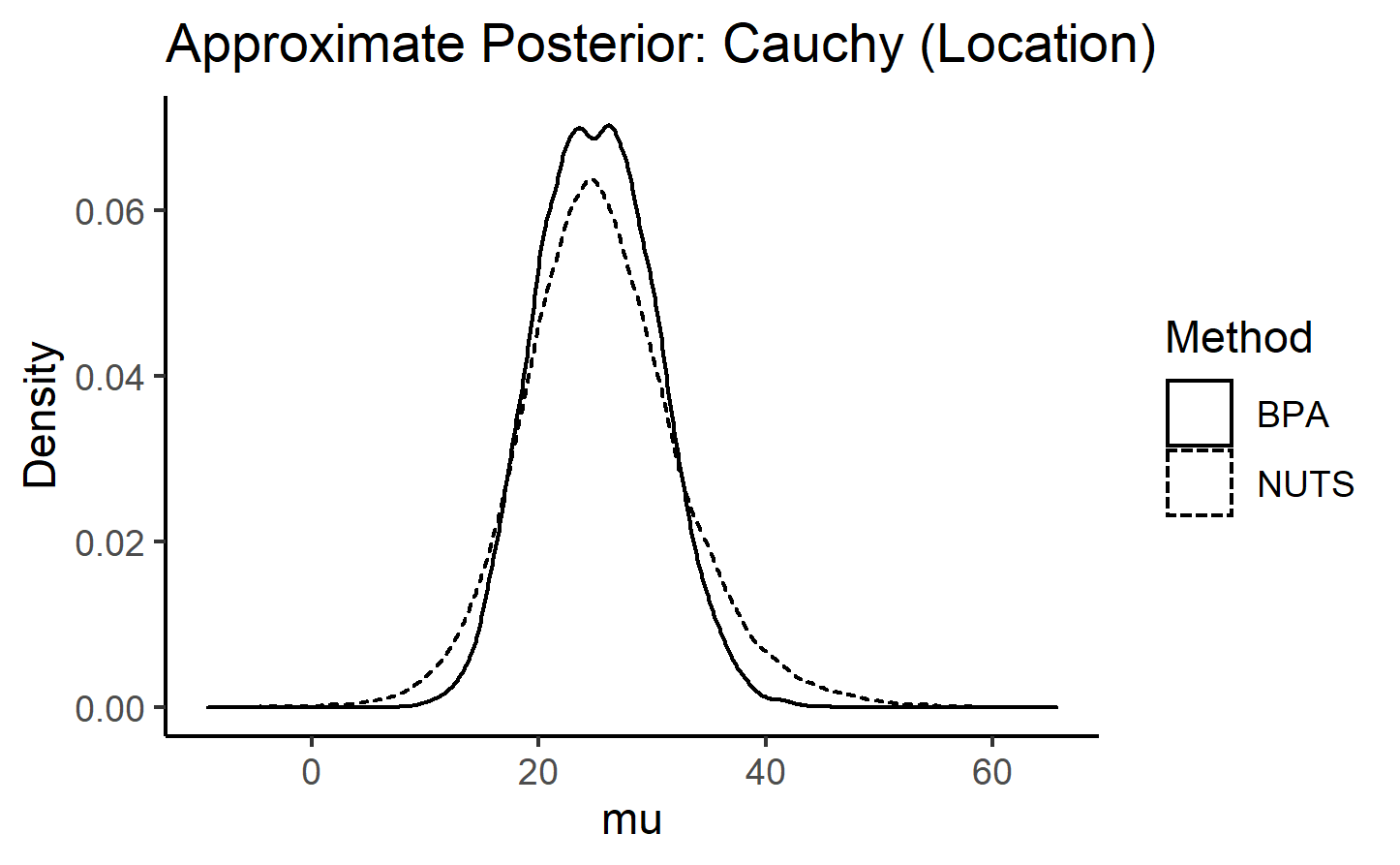}
    \caption{The samples from the posterior distribution from the BPA and NUTS method for the non-informative prior $\displaystyle P(\mu,\sigma) \propto \sigma^{-1}$. The total number of post warm-up samples for both methods is 50000.}
    \label{fig:cauchy_mean}
\end{figure}

\begin{figure}[!ht]
    \centering
    \includegraphics[width=3in]{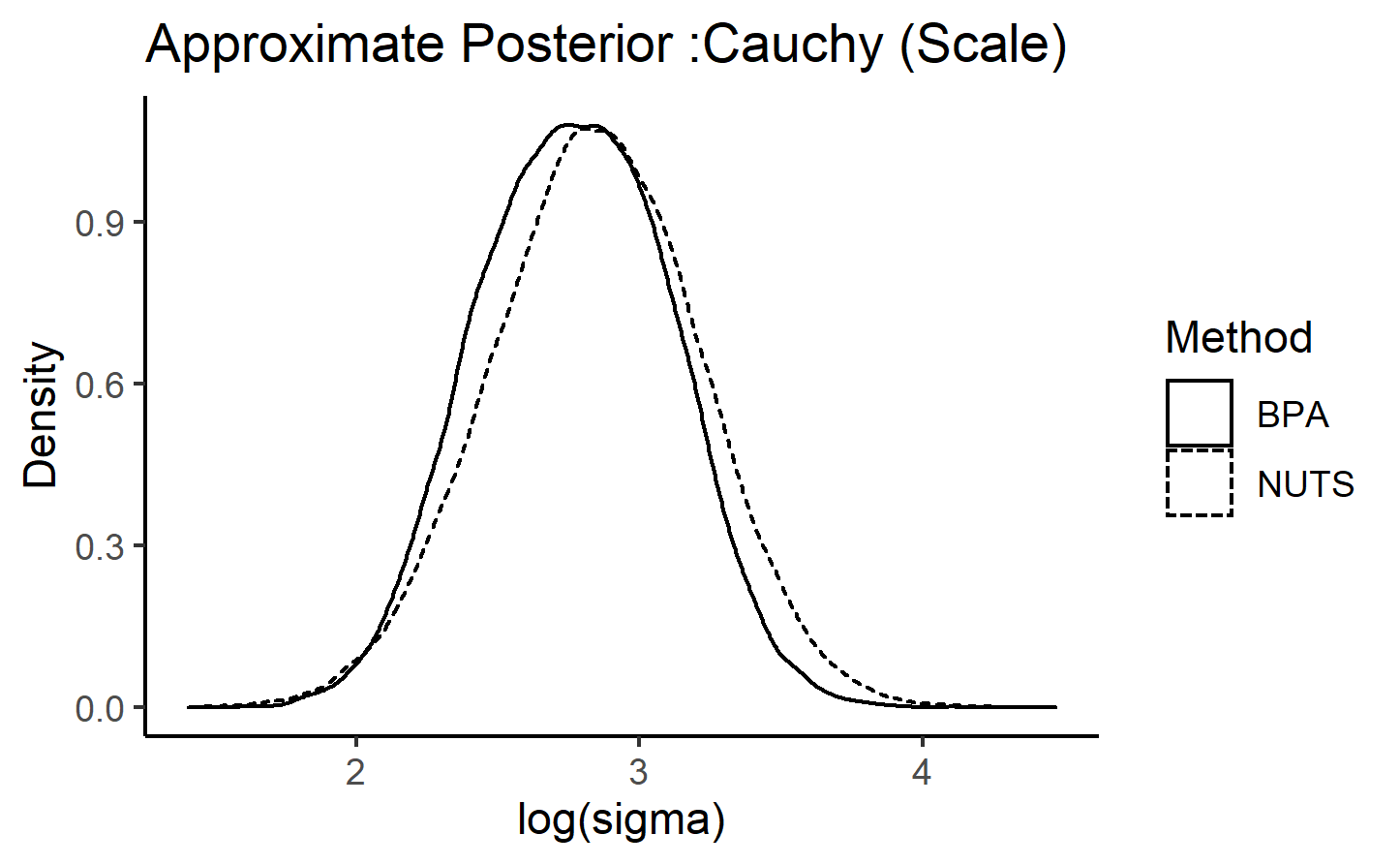}
    \caption{The samples from the posterior distribution from the BPA and NUTS method for the non-informative prior $\displaystyle P(\mu,\sigma) \propto \sigma^{-1}$. The total number of post warm-up samples for both methods is 50000.}
    \label{fig:cauchy_sd}
\end{figure}

For the multiparameter model, the second-order partial differential equations from the Euler-Lagrange equations were set up and solved numerically. The joint prior used for this analysis is the non-informative prior $\displaystyle P(\mu,\sigma) \propto \sigma^{-1}$. Both algorithms used 100 balls/chains with a chain length of 1000 for their respective implementations. The resulting approximate posterior distributions from both methods are shown in Figures \ref{fig:cauchy_mean} and \ref{fig:cauchy_sd}.

We see that there was a good correspondence between the shape of the posterior distributions for both $\mu$ and $\log \sigma$, with the BPA yielding slightly narrower results compared to the NUTS algorithm. Table \ref{tab:cauchy} also shows the resulting summary statistics for the two parameters from the two methods.

\begin{table}[H]
\caption{Estimates of the posterior distribution from NUTS and the Ball Pit Algorithm (BPA) for the multi-parameter Cauchy likelihood model.} 
\label{tab:cauchy}
\begin{adjustbox}{width=1\textwidth}
\begin{tabular}{|c|c|c|c|c|c|c|c|c|c|}
\hline
\textbf{Parameter}                      & Method        & \textbf{Mean} & \textbf{Std. Error.} & \textbf{Std. Dev.} & \textbf{2.5\%} & \textbf{25\%} & \textbf{50\%} & \textbf{75\%} & \textbf{97.5\%} \\ \hline
\multirow{2}{*}{\textbf{$\mu$}}         & \textbf{STAN} & 25.54         & 0.04                 & 7.05               & 12.58          & 20.96         & 25.12         & 29.67         & 40.96           \\ \cline{2-10} 
                                        & \textbf{BPA}  & 25.1          & 0.45                & 5.20               & 15.50          & 21.34         & 25.02         & 28.71         & 35.44           \\ \hline
\multirow{2}{*}{\textbf{$\log \sigma$}} & \textbf{STAN} & 2.84          & \textless{}0.01      & 0.37               & 2.11           & 2.60          & 2.85          & 3.10          & 3.57            \\ \cline{2-10} 
    & \textbf{BPA}  & 2.84          & 0.10               & 0.66              & 2.00           & 2.50          & 2.79          & 3.06          & 3.79            \\ \hline
\end{tabular}
\end{adjustbox}
\end{table}

The computational time of the BPA to sample from the marginal posterior distributions of both $\mu$ and $\log \sigma$ was measured to be 22.08 seconds, which was 55\% lower than the computational time of NUTS.

We note that there was a bias of 0.4 between the posterior mean estimates for $\mu$ obtained from STAN and the BPA. Although this bias is small compared to its magnitude (2\% of the mean), this difference still needs to be accounted for. This deviation can be explained by the arbitrary nature of the limits of the uniform distribution as the prior. Simulating from a uniform distribution with a different width affects the mean and the quantile estimates because of the dynamics of the balls from points far away from the expected posterior value. For this example, we used starting points simulated from $Unif(0,100)$, but changing the upper limit to 50 or 1000 gave us different results. The approximate posterior obtained from the BPA is slightly narrower compared to the resulting posterior from NUTS. However, the estimated posterior median  from the BPA was consistent with the other estimation methods such as the Random Walk Metropolis-Hastings MCMC, Independence sampling, and Gibbs Sampling as calculated by \cite{albert_bayesian_2009}.

The posterior distribution for the logarithm of the scale parameter, which we refer to as the log-scale parameter, obtained from the BPA is similar to the resulting posterior distribution from NUTS as shown in Figure \ref{fig:cauchy_sd}. We also observed a slightly heavier tail for the BPA, which after further investigation was determined to be caused by the truncation set when we simulated from the uniform prior. To account for this, we investigated the posterior median instead of the posterior mean in comparing the two methods. The posterior medians suggest that the BPA is shifted to the left compared to the NUTS distribution. Based on the values provided by \cite{albert_bayesian_2009}, the posterior median resulting from NUTS was close to other Monte Carlo methods (2.85) and other samplers such as the Gibbs sampler (2.86) while the posterior median for the logarithm of the scale parameter obtained by the BPA (2.79) is close to the same posterior median obtained when a Laplace normal approximation (2.77) is used \citep{albert_bayesian_2009}. However, the 95\% credible intervals are wider for the BPA approximation compared to the Laplace normal approximation. 

\subsection{Discussion}

The resulting distributions from the BPA are comparable to the distributions obtained from the HMC implementation through the NUTS algorithm; however, a notable difference between the results of the BPA and NUTS was that the BPA generally yielded a narrower distribution as attested by the lower standard deviations and quantile estimates. This trend was observed for distributions with unbounded or half-bounded supports for its parameters such as the Poisson distribution ($ \lambda \in [0,\infty)$) and the Cauchy distribution ($\mu \in (-\infty,\infty)$ and $\sigma \in [0,\infty)$). It is imperative to know where to truncate the improper priors when sampling for initial points using the BPA since the resulting distributions are affected by where the chain starting points are located. If the chosen starting point is far away from the posterior mode, the chain might not converge. Aside from the nature of the parameter supports, the choice of prior can also lead to narrower posterior distributions after implementing the BPA. If the starting points are confined within a narrow region of the parameter space, it might not be able to explore regions that was covered by the HMC. 

The BPA also reduced the computational time by at least 95\% compared to the \texttt{rstan} implementation for the single parameter models and 55\% for the Cauchy model. As stated in Section \ref{subsec:adv}, the HMC entails simulating momentum values at the start of each iteration to keep the value of the Hamiltonian constant. With the BPA, there is no need to simulate velocity values after each iteration since the trajectory is dictated by the probability of taking a certain path given by the path integral formulation. Aside from the rejection of the candidate point, the only instances where the velocities are resampled are when the chain is stuck or the likelihood at the candidate point is undefined. The fast convergence of the BPA also allows us to use shorter chains which also leads to faster sampling from the target distribution. This significant decrease in computational time shows the potential of the BPA for sampling target distributions more efficiently compared to HMC and other standard MCMC algorithms.

\section{Conclusion and Next Steps}

Using path integrals in the Bayesian analysis of Markov chains provided valuable information on how the prior distribution evolves to the posterior distribution. We found that the path followed by the Markov chain in the parameter space is dictated by the value of the log-likelihood and the noise strength of the Markov process. Using this information, we devised the BPA, which is a Monte Carlo algorithm that utilizes a Lagrangian framework to predict the trajectory of a Markov chain in the parameter space. We then compared the BPA to the Hamiltonian Monte Carlo, which is another standard sampling method that uses concepts from physical systems, as implemented by the NUTS algorithm in the R package \texttt{rstan}. The BPA was found to be at least 95\% faster than the NUTS algorithm in sampling posterior distributions for single parameter models and 55\% faster in sampling the marginal posterior distributions for the Cauchy likelihood parameters. The resulting approximate posterior distributions from the BPA is comparable to the ones generated using NUTS, which is supported by visual and numerical evidences. Furthermore, the standard deviation for the BPA posterior distributions are generally lower than the NUTS posterior distributions.

The BPA was also applied to a multi-parameter Cauchy model to estimate the location and the logarithm of the scale parameters. A slight bias was observed for the estimated posterior means for the location parameter, but the posterior median of the location parameter was found to be consistent between other MCMC methods. As for the log-scale parameter, the estimated posterior median from the result of the BPA is close to the posterior median when the Laplace normal approximation is used. The distribution appears to be slightly shifted to the left and is slightly wider than the posterior distribution obtained from NUTS.

We aim to extend the BPA to sample from joint distributions for multi-parameter models by accounting for multiple Markovian processes in the parameter space. Sampling from the joint distribution of the parameters would only need one implementation of the BPA instead of its current state where it needs multiple implementations to sample from marginal distributions.  We believe that this will improve the performance of the algorithm in approximating the exact posterior distributions of multi-parameter models.

\bibliographystyle{jabes}

\bibliography{main.bib}

\begin{thebibliography}{38}
\newcommand{\enquote}[1]{``#1''}
\expandafter\ifx\csname natexlab\endcsname\relax\def\natexlab#1{#1}\fi
\expandafter\ifx\csname url\endcsname\relax
  \def\url#1{\texttt{#1}}\fi
\expandafter\ifx\csname urlprefix\endcsname\relax\def\urlprefix{URL }\fi

\bibitem[{Albert(2009)}]{albert_bayesian_2009}
Albert, J. (2009), \textit{Bayesian computation with {R}}, Springer Science \&
  Business Media.

\bibitem[{Albert(2018)}]{learnbayes2018}
--- (2018), \textit{LearnBayes: Functions for Learning Bayesian Inference},
  \urlprefix\url{https://CRAN.R-project.org/package=LearnBayes}. R package
  version 2.15.1.

\bibitem[{Albeverio et~al.(1976)Albeverio, H{\"o}egh-Krohn, and
  Mazzucchi}]{albeverio1976mathematical}
Albeverio, S., H{\"o}egh-Krohn, R., and Mazzucchi, S. (1976),
  \textit{Mathematical theory of Feynman path integrals}, vol. 523, Springer.

\bibitem[{Andrieu et~al.(2009)Andrieu, Doucet, and
  Holenstein}]{andrieu2009particle}
Andrieu, C., Doucet, A., and Holenstein, R. (2009), \enquote{Particle markov
  chain monte carlo for efficient numerical simulation,} in \textit{Monte Carlo
  and quasi-Monte Carlo methods 2008}, Springer, pp. 45--60.

\bibitem[{Andrieu et~al.(2010)Andrieu, Doucet, and
  Holenstein}]{andrieu2010particle}
--- (2010), \enquote{Particle markov chain monte carlo methods,}
  \textit{Journal of the Royal Statistical Society: Series B (Statistical
  Methodology)}, 72, 269--342.

\bibitem[{Betancourt(2017)}]{betancourt2017conceptual}
Betancourt, M. (2017), \enquote{A conceptual introduction to hamiltonian monte
  carlo,} \textit{arXiv preprint arXiv:1701.02434}.

\bibitem[{Caldeira and Leggett(1983)}]{caldeira1983path}
Caldeira, A.~O. and Leggett, A.~J. (1983), \enquote{Path integral approach to
  quantum brownian motion,} \textit{Physica A: Statistical mechanics and its
  Applications}, 121, 587--616.

\bibitem[{Casella and George(1992)}]{casella1992explaining}
Casella, G. and George, E.~I. (1992), \enquote{Explaining the gibbs sampler,}
  \textit{The American Statistician}, 46, 167--174.

\bibitem[{Chang et~al.(2015)Chang, Fok, and Chou}]{chang2015bayesian}
Chang, J.~C., Fok, P.-W., and Chou, T. (2015), \enquote{Bayesian uncertainty
  quantification for bond energies and mobilities using path integral
  analysis,} \textit{Biophysical journal}, 109, 966--974.

\bibitem[{Chang et~al.(2014)Chang, Savage, and Chou}]{chang2014path}
Chang, J.~C., Savage, V.~M., and Chou, T. (2014), \enquote{A path-integral
  approach to bayesian inference for inverse problems using the semiclassical
  approximation,} \textit{Journal of Statistical Physics}, 157, 582--602.

\bibitem[{Chen et~al.(2003)}]{chen2003bayesian}
Chen, Z. et~al. (2003), \enquote{Bayesian filtering: From kalman filters to
  particle filters, and beyond,} \textit{Statistics}, 182, 1--69.

\bibitem[{Del~Moral(1996)}]{del1996nonlinear}
Del~Moral, P. (1996), \enquote{Nonlinear filtering using random particles,}
  \textit{Theory of Probability \& Its Applications}, 40, 690--701.

\bibitem[{Doucet et~al.(2001)Doucet, De~Freitas, and
  Gordon}]{doucet2001introduction}
Doucet, A., De~Freitas, N., and Gordon, N. (2001), \enquote{An introduction to
  sequential monte carlo methods,} in \textit{Sequential Monte Carlo methods in
  practice}, Springer, pp. 3--14.

\bibitem[{Duane et~al.(1987)Duane, Kennedy, Pendleton, and
  Roweth}]{duane1987hybrid}
Duane, S., Kennedy, A.~D., Pendleton, B.~J., and Roweth, D. (1987),
  \enquote{Hybrid monte carlo,} \textit{Physics letters B}, 195, 216--222.

\bibitem[{Feynman et~al.(2010)Feynman, Hibbs, and Styer}]{feynman2010quantum}
Feynman, R.~P., Hibbs, A.~R., and Styer, D.~F. (2010), \textit{Quantum
  mechanics and path integrals}, Courier Corporation.

\bibitem[{Fujii and Hatakenaka(2018)}]{fujii2017path}
Fujii, T. and Hatakenaka, N. (2018), \enquote{A path integral approach to
  bayesian inference in markov processes,} \textit{Global Journal of Pure and
  Applied Mathematics}, 14, 721--731.

\bibitem[{Gelman et~al.(2013)Gelman, Carlin, Stern, Dunson, Vehtari, and
  Rubin}]{gelman2013bayesian}
Gelman, A., Carlin, J.~B., Stern, H.~S., Dunson, D.~B., Vehtari, A., and Rubin,
  D.~B. (2013), \textit{Bayesian data analysis}, CRC press.

\bibitem[{Geman and Geman(1984)}]{geman1984stochastic}
Geman, S. and Geman, D. (1984), \enquote{Stochastic relaxation, gibbs
  distributions, and the bayesian restoration of images,} \textit{IEEE
  Transactions on pattern analysis and machine intelligence}, 721--741.

\bibitem[{Girolami and Calderhead(2011)}]{girolami2011riemann}
Girolami, M. and Calderhead, B. (2011), \enquote{Riemann manifold langevin and
  hamiltonian monte carlo methods,} \textit{Journal of the Royal Statistical
  Society: Series B (Statistical Methodology)}, 73, 123--214.

\bibitem[{Glimm and Jaffe(2012)}]{glimm2012quantum}
Glimm, J. and Jaffe, A. (2012), \textit{Quantum physics: a functional integral
  point of view}, Springer Science \& Business Media.

\bibitem[{Goldstein et~al.(2002)Goldstein, Poole, and
  Safko}]{goldstein2002classical}
Goldstein, H., Poole, C., and Safko, J. (2002), \enquote{Classical mechanics,}
  .

\bibitem[{Hawking(1979)}]{hawking1979path}
Hawking, S.~W. (1979), \enquote{The path-integral approach to quantum gravity,}
  in \textit{General relativity}.

\bibitem[{Hoffman et~al.(2014)Hoffman, Gelman et~al.}]{hoffman2014no}
Hoffman, M.~D., Gelman, A., et~al. (2014), \enquote{The no-u-turn sampler:
  adaptively setting path lengths in hamiltonian monte carlo.} \textit{J. Mach.
  Learn. Res.}, 15, 1593--1623.

\bibitem[{Holenstein(2009)}]{holenstein2009particle}
Holenstein, R. (2009), \enquote{Particle markov chain monte carlo,} PhD thesis,
  University of British Columbia.

\bibitem[{Lan et~al.(2015)Lan, Stathopoulos, Shahbaba, and
  Girolami}]{lan2015markov}
Lan, S., Stathopoulos, V., Shahbaba, B., and Girolami, M. (2015),
  \enquote{Markov chain monte carlo from lagrangian dynamics,} \textit{Journal
  of Computational and Graphical Statistics}, 24, 357--378.

\bibitem[{Liu and Chen(1998)}]{liu1998sequential}
Liu, J.~S. and Chen, R. (1998), \enquote{Sequential monte carlo methods for
  dynamic systems,} \textit{Journal of the American statistical association},
  93, 1032--1044.

\bibitem[{MacKenzie(2000)}]{mackenzie2000path}
MacKenzie, R. (2000), \enquote{Path integral methods and applications,}
  \textit{arXiv preprint quant-ph/0004090}.

\bibitem[{Pompe et~al.(2020)Pompe, Holmes, and
  {\L}atuszy{\'n}ski}]{pompe2020framework}
Pompe, E., Holmes, C., and {\L}atuszy{\'n}ski, K. (2020), \enquote{A framework
  for adaptive mcmc targeting multimodal distributions,} \textit{The Annals of
  Statistics}, 48, 2930--2952.

\bibitem[{{R Core Team}(2017)}]{rcite}
{R Core Team} (2017), \textit{R: A Language and Environment for Statistical
  Computing}, R Foundation for Statistical Computing, Vienna, Austria,
  \urlprefix\url{https://www.R-project.org/}.

\bibitem[{Robert(2007)}]{robert2007bayesian}
Robert, C. (2007), \textit{The Bayesian choice: from decision-theoretic
  foundations to computational implementation}, Springer Science \& Business
  Media.

\bibitem[{Sakurai and Napolitano(2011)}]{sakurai2011modern}
Sakurai, J. and Napolitano, J. (2011), \enquote{Modern quantum mechanics 2nd
  ed,} .

\bibitem[{Smith and Roberts(1993)}]{smith1993bayesian}
Smith, A.~F. and Roberts, G.~O. (1993), \enquote{Bayesian computation via the
  gibbs sampler and related markov chain monte carlo methods,} \textit{Journal
  of the Royal Statistical Society: Series B (Methodological)}, 55, 3--23.

\bibitem[{{Stan Development Team}(2020)}]{rstan}
{Stan Development Team} (2020), \enquote{{RStan}: the {R} interface to {Stan},}
  \urlprefix\url{http://mc-stan.org/}. R package version 2.21.2.

\bibitem[{Szabados(2010)}]{szabados_elementary_2010}
Szabados, T. (2010), \enquote{An elementary introduction to the {Wiener}
  process and stochastic integrals,} \textit{arXiv:1008.1510 [math]},
  \urlprefix\url{http://arxiv.org/abs/1008.1510}. ArXiv: 1008.1510.

\bibitem[{Tipler and Mosca(2007)}]{tipler2007physics}
Tipler, P.~A. and Mosca, G. (2007), \textit{Physics for scientists and
  engineers}, Macmillan.

\bibitem[{Wiegel(1975)}]{wiegel1975path}
Wiegel, F.~W. (1975), \enquote{Path integral methods in statistical mechanics,}
  \textit{Physics Reports}, 16, 57--114.

\bibitem[{Wio(2013)}]{wio2013path}
Wio, H.~S. (2013), \textit{Path integrals for stochastic processes: An
  introduction}, World Scientific.

\bibitem[{Yuan et~al.(2012)Yuan, Girolami, and Niranjan}]{yuan2012markov}
Yuan, K., Girolami, M., and Niranjan, M. (2012), \enquote{Markov chain monte
  carlo methods for state-space models with point process observations,}
  \textit{Neural Computation}, 24, 1462--1486.

\end{thebibliography}

\end{document}